\documentclass[conference]{IEEEtran}

\usepackage{floatflt}
\usepackage{color}
\usepackage{cite}
\usepackage{graphicx}
\usepackage{amsmath}
\usepackage{amsthm}
\usepackage{amssymb}
\usepackage{verbatim}
\usepackage{psfrag,array}
\usepackage{subfigure}
\usepackage{multirow}
\usepackage{hhline}
\usepackage{stfloats}
\usepackage{amsmath}
\usepackage{epstopdf}
\usepackage{algorithm}
\usepackage{algpseudocode}
\usepackage{multirow}

\newcommand{\p}[1]{\mathop{\mbox{\it p} } }

\renewcommand{\vec}[1]{\ensuremath{\boldsymbol{#1}}}
\newcommand{\be}{\begin{equation}}
\newcommand{\ee}{\end{equation}}
\newcommand{\ba}{\begin{array}}
\newcommand{\ea}{\end{array}}
\newcommand{\bea}{\begin{eqnarray}}
\newcommand{\eea}{\end{eqnarray}}
\newcommand{\bean}{\begin{eqnarray*}}
\newcommand{\eean}{\end{eqnarray*}}
\newcommand{\argmax}{\mathop{\arg\max}}

\newcommand{\rmh}{^{\dag}}

\definecolor{white}{rgb}{1,1,1}

\newtheorem{lemma}{Lemma}

\newtheorem{proposition}{Proposition}

\begin{document}

\title{Unmanned Aerial Vehicle Assisted Cellular Communication}
\author
{
Sha Hu, Jose Flordelis, Fredrik Rusek, and Ove Edfors\\
Department of Electrical and Information Technology, \\Lund University, Lund, Sweden \\
email: \{firstname.lastname\}@eit.lth.se
\vspace{-3mm}
}

\maketitle

\begin{abstract}
In this paper, we consider unmanned aerial vehicle (UAV) assisted cellular communications where UAVs are used as amplify-and-forward (AF) relays. The effective channel with UAV-assisted communication is modeled as a Rayleigh product channel, for which we derive a tight lower-bound of the ergodic capacity in closed-form. With the obtained lower-bound, trade-offs between the transmit power and the equipped number of antennas of the UAVs can be analyzed. Alternatively, for a given setting of users and the base-transceiver station (BTS), the needed transmit power and number of antennas for the UAVs can be derived in order to have a higher ergodic capacity with the UAV-assisted communication than that without it.
\end{abstract}

\section{Introduction}
Due to a rapidly growing market, unmanned aerial vehicles (UAVs) have recently gained attentions in many applications. UAVs can be used in cellular and satellite communication systems to improve data connections between a base-transceiver station (BTS) and users that are far from the BTS or obstructed by surrounding objects such as tall buildings and mountains~\cite{CG17, CE17, ZL16, WZ18, HB15}. UAVs can improve data transmission in various ways. Firstly, due to its height in the air, a UAV can have line-of-sight (LoS) to the BTS which increases the received signal-to-noise ratio (SNR). Secondly, a UAV can use a higher transmit power with equipped large-capacity battery or with solar-charging systems~\cite{MP15}. Thirdly, a UAV can easily adjust its gesture in the sky to beamform the relayed data into a better direction to the BTS. Lastly, UAVs can appear anywhere when there is a need which yields flexible and low-cost network deployments.

A typical scenario of UAV-assisted cellular communication system is depicted in Fig.~1. In a simple form, UAVs can be used as amplify-and-forward (AF) relays~\cite{GV97} to assist users when they are at cell edges or in deep shadow fading. The UAVs can also be more advanced such as with capabilities of beamforming with gesture adjustments and digital precoding. To simplify the analysis, we model the UAV-assisted cellular communication system as a Rayleigh product model when UAVs are beyond LoS. The Rayleigh product channel arises from a general double-scattering model~\cite{GP02}, and has been considered before in other contexts e.g.,~\cite{FM10, YB07, JG08, XS10, BJ06, LH05}. However, these works are more focused on analyzing the statistics of eigenvalues and in general have complex expressions for the ergodic capacity. 

In this paper, we take a special interest in comparing the ergodic capacity between the UAV-assisted communication and the one without it. We derive a lower-bound for the ergodic capacity of the UAV-assisted model in closed-form, which is shown to be tight. The lower-bound provides insights about the trade-off between the transmit power and the number of antennas needed of the UAVs. It is also helpful to aid in the designs of UAV-assisted cellular networks, for tasks such as specifying the number of antennas and transmit power of the UAVs to achieve certain ergodic capacity, or maximizing the utility of each spent-antenna with a given transmit power.

\begin{figure}[t]
\vspace*{-0mm}
\begin{center}
\hspace*{4mm}
\scalebox{0.31}{\includegraphics{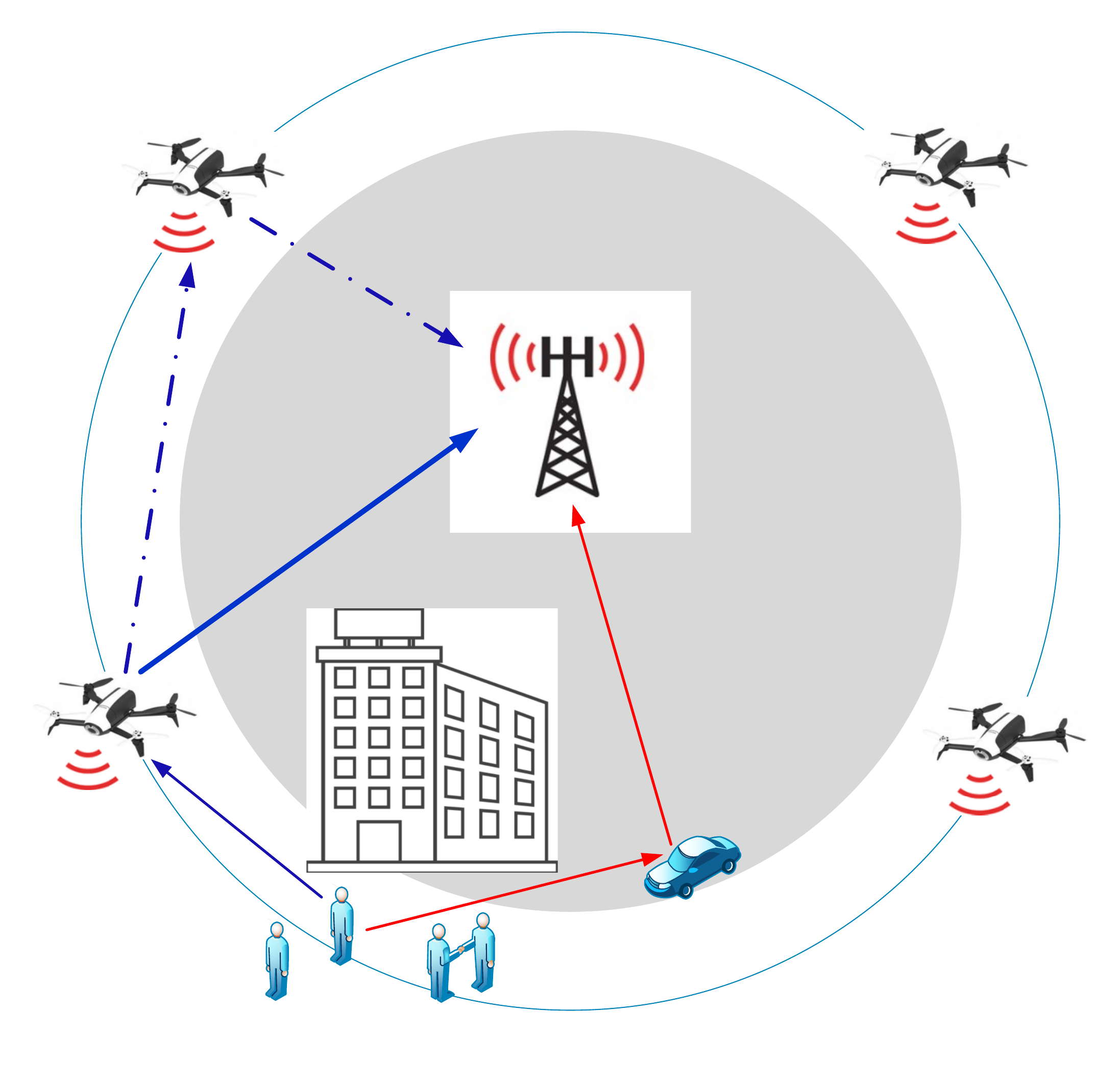}}
\vspace*{-6mm}
\caption{\label{fig1}UAV-assisted cellular communication when users are in deep shadow fading and at cell edge, where UAVs are used as AF relays. }
\vspace*{-7mm}
\end{center}
\end{figure}

\section{Capacity with UAV-Assisted Communication}

Let\rq{}s consider two different approaches for users connecting to a BTS. The first one is that users, with a total number of $M$ transmit antennas, are directly communicating with the BTS, which is equipped with $N$ receive antennas. The received signal at BTS reads
\bea \label{md1} \vec{y}=\sqrt{p}\vec{H}\vec{x}+\vec{n}, \eea
where $\vec{x}$ comprises the transmitted signal from one or multiple users. The Rayleigh multi-input multi-output (MIMO) channel $\vec{H}$ is of size $N\!\times\!M$ and comprises independent and identically distributed (i.i.d.) complex-valued Gaussian elements with zero-mean and unit-variance. For simplicity, we assume that $\vec{n}$ is additive white Gaussian noise (AWGN) with an identity covariance matrix. We let $p$ denote the transmit-power\footnote{In this paper, we abuse the term \lq\lq{}transmit power\rq\rq{} by including distance-dependent power attenuation and large-scale fading.} on each user antenna.

The capacity in this case equals
\bea \label{R1} R=\log\det\!\left(\vec{I}+p\vec{H}\rmh\vec{H}\right)\!.\eea
When users are in deep fading or at cell edges, they suffer from poor data connections due to a low received SNR at the BTS, which equals $\mathrm{tr}(\mathbb{E}[p\vec{H}\vec{H}\rmh]) \!=\! pMN$. Increasing transmit power has several disadvantages as it raises interference to neighboring users and also consumes more battery-power. In extreme situations such as natural hazards, the battery-life is important for users to maintain long-term connections to the rescuers. On the other hand, UAVs can be used as AF relays to improve data-transmission of the users through at least two possible means: amplify the received signal from the users with a higher transmit power; and redirect the signal into the direction of the BTS.

Assuming such a UAV-assisted cellular communication scenario, where the UAVs are equipped with a total $K$ transmit and receive antennas\footnote{The $K$ antennas can belong to a a single UAV or multiple UAVs.}, the received signal in this case yields a Rayleigh product channel model
 \bea \label{md2}  \vec{y}= \sqrt{q}\vec{Q}\vec{x}+\vec{n}, \eea
where $\vec{Q}\!=\!\vec{Q}_2\vec{Q}_1$, and $\vec{Q_1}$ and $\vec{Q}_2$ are the MIMO channels from users to the UAV with size $K\!\times\!M$, and from the UAV to the BTS with size $N\!\times\!K$, respectively. Similarly to (\ref{md1}), we denote $q$ as the transmit power and model $\vec{Q_1}$ and $\vec{Q}_2$ as Rayleigh channels that comprise i.i.d. complex-valued Gaussian elements with zero-mean and unit-variance, and the noise $\vec{n}$ is the same as in (\ref{md1}).

The capacity corresponding to (\ref{md2}) equals
\bea \label{S1} S=\frac{1}{1+\tau}\log\det\!\left(\vec{I}+q\vec{Q}\rmh\vec{Q}\right)\!, \eea
where $\tau$ denotes the additional time-delay in UAV-assisted transmissions \cite{FM10}. In a pipelined scheme $\tau$ can be negligible \cite{NK04}, and we let $\tau\!=\!0$ in the discussions.

The received SNR at the BTS in this case equals $\mathrm{tr}(\mathbb{E}[q\vec{Q}\vec{Q}\rmh]) \!=\! qMKN$. To have a higher SNR than the case with direct transmission, it requires
\bea \label{cond1} qK\!>\!p. \eea

Instead of the received SNR, we are also interested in comparing the ergodic capacities in these two cases. Especially when $K\!<\!M$, that is, the number of antennas equipped with the UAVs is less than that of the users. In this case, the spatial multiplexing gain is reduced. This can be due to a large number of users in difficult situations at the same time, or the UAVs have low-cost designs with limit numbers of antennas. According to (\ref{cond1}), when $K$ is small the transmit power $q$ has to increase. But as UAVs use built-in battery, the power-capacity can also be limited. Therefore, it is of interest to evaluate the ergodic capacity in relation to parameters $q$ and $K$ for the UAV-assisted systems, and understand when it is beneficial to use UAVs for assistances.

\section{A Lower-Bound on Ergodic Capacity}
In this section we derive a lower-bound for the Rayleigh product channel model (\ref{md2}). A similar analysis can be carried out for multi-tier connections through UAVs, that is, the product channel comprises more than two components. At the begining we assume $N\!\geq\! M, K$, but as it will become clear later, such an assumption is not needed for the validity of the derived lower-bound.

\subsection{The Case $K\!\leq\!M$ }

For the purpose of comparison, we first find an upper-bound for the direct communication between the users and the BTS. By Jensen\rq{}s inequality, the ergodic capacity $\tilde{R}\!=\!\mathbb{E}[R]$ corresponding to the direct approach (\ref{R1}) is upper bounded as
{\setlength\arraycolsep{2pt} \bea \label{R3} \tilde{R}&\leq&\log\det\!\left(\vec{I}+p\mathbb{E}[\vec{H}\rmh\vec{H}]\right)\notag \\
&=&M\log\!\left(1+pN\right ),\eea}
\hspace{-2.02mm}which is tight when the number of receiver antennas $N$ is large such as with massive MIMO systems \cite{T99}.

To derive a lower-bound for the UAV-assisted case, we fist note that
\bea \det\!\left(\vec{I}+q\vec{Q}\rmh\vec{Q}\right)=\det\!\left(\vec{I}+q\vec{\Sigma}_1\vec{\Sigma}_2\right)\!, \notag \eea
where $\vec{\Sigma}_1\!=\!\vec{Q}_1\vec{Q}_1\rmh$ and $\vec{\Sigma}_2\!=\!\vec{Q}_2\rmh\vec{Q}_2$ are $K\!\times\!K$ matrices. 

Using Minkowski\rq{}s inequality
\bea  \det(\vec{A}+\vec{B})^{1/K}\geq \det(\vec{A})^{1/K}+\det(\vec{B})^{1/K}, \notag \eea
the ergodic capacity $\tilde{S}\!=\!\mathbb{E}[S]$, corresponding to the UAV-assisted approach (\ref{S1}), satisfies
{\setlength\arraycolsep{2pt} \bea  \tilde{S}&\geq&K\mathbb{E}\!\left[\log\!\left(1+q\big(\!\det(\vec{\Sigma}_1\vec{\Sigma}_2)\big)^{1/K}\right)\!\right]\notag \\
&=&K\mathbb{E}\!\left[\log\!\left(1+q\exp\!\left(\frac{1}{K}\ln\det(\vec{\Sigma}_1\vec{\Sigma}_2)\right)\!\right)\!\right]\!. \notag\eea}
\hspace{-1.5mm}Again by Jensen\rq{}s inequality, it holds that
\bea \label{S3} \tilde{S}\geq K\log\!\left(1+q\exp\!\left(\frac{1}{K}\mathbb{E}\!\left[\ln\det(\vec{\Sigma}_1\vec{\Sigma}_2)\right]\right)\!\right)\!.\eea
Since
 \bea \label{c1} \ln\det(\vec{\Sigma}_1\vec{\Sigma}_2)=\ln\det(\vec{\Sigma}_1)+\ln\det(\vec{\Sigma}_2), \eea
where $\vec{\Sigma}_1$, and $\vec{\Sigma}_2$ are complex Wishart distributed, it is readily seen from \cite{OP02, G63} that
 \bea \label{c2} \mathbb{E}\!\left[\ln\det(\vec{\Sigma}_1)\right]\!\!\!&=&\!\!\!\sum_{\ell=1}^{K}\psi(M\!-\!\ell\!+\!1),  \\
 \label{c3} \mathbb{E}\!\left[\ln\det(\vec{\Sigma}_2)\right]\!\!\!&=&\!\!\!\sum_{\ell=1}^{K}\psi(N\!-\!\ell\!+\!1),   \eea
where $\psi(n)\!=\!-\gamma\!+\!\sum\limits_{k=1}^{n-1}\frac{1}{k}$ is the \textit{digamma} function \cite{ET55} and $\gamma\!\approx\!0.5772$ is the \textit{Euler-Mascheroni} constant. 

Combining (\ref{c1})-(\ref{c3}) yields
\bea \frac{1}{K}\mathbb{E}\left[ \ln\det\!\left(\vec{\Sigma}_1\vec{\Sigma}_2\right)\right]=g(K)-2\gamma, \notag \eea
where
 \bea g(K)=\frac{1}{K}\sum_{\ell=1}^{K}\left(\sum_{m=1}^{M-\ell}\frac{1}{m}+\sum_{n=1}^{N-\ell}\frac{1}{n}\right)\!. \notag\eea
Hence, from (\ref{S3}) the ergodic capacity $\tilde{S}$ for the UAV-assisted communication is lower bounded as
 \bea \label{SLB} \tilde{S}\geq K\log\!\left(1+q\exp\!\big(g(K)-2\gamma\big)\right)\!.\eea

\begin{figure}[t]
\vspace{-3mm}
\begin{center}
\hspace*{-2mm}
\scalebox{0.47}{\includegraphics{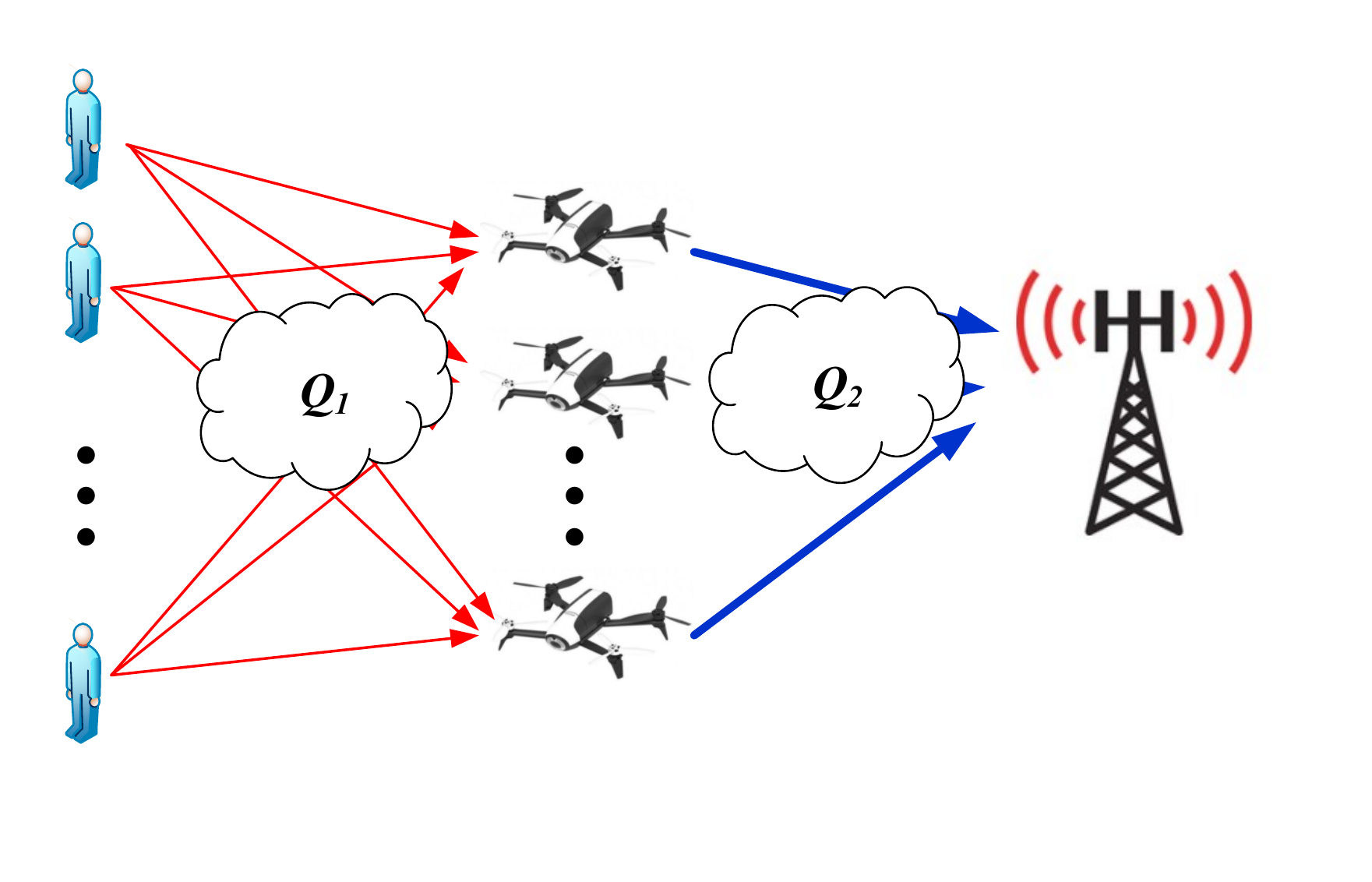}}
\vspace{-11mm}
\caption{\label{fig2} The Rayleigh fading product channel with UAV-assisted communication that is specified by three parameters $(M, K, N)$ that are the number of antennas of users, UAVs, and BTS, respectively.}
\vspace{-7mm}
\end{center}
\end{figure}

\subsection{The Case $K\!>\!M$}
Although $K\!\leq\!M$ is more interesting, we next also consider the case $K\!>\!M$. Denote the singular-value decomposition (SVD)
\bea \label{svd1}  \vec{Q}_1\vec{Q}_1\rmh= \vec{U}\rmh\vec{\Lambda}_1\vec{U}, \eea
where $K\!\times\!K$ matrices $\vec{U}$ is unitary, and $\vec{\Lambda}_1$ is diagonal with the last $K\!-\!M$ diagonal elements being 0s. 

Then, it holds that
{\setlength\arraycolsep{0pt} \bea \label{switch}  &&\mathbb{E}_{\{\vec{Q}_1,\,\vec{Q}_2\}}\!\!\left[\log\det\!\left(\vec{I}+q\vec{Q}\rmh\vec{Q}\right)\!\right] \notag \\
&&=\!\mathbb{E}_{\{\vec{Q}_2,\,\vec{U},\,\vec{\Lambda}_1\}}\!\!\left[\log\det\!\left(\vec{I}+q\vec{Q}_2 \vec{U}\rmh\vec{\Lambda}_1\vec{U}\vec{Q}_2\rmh\right)\!\right] \notag \\
&&=\!\mathbb{E}_{\{\tilde{\vec{Q}}_2=\vec{Q}_2 \vec{U}\rmh,\,\vec{\Lambda}_1\}}\!\!\left[\log\det\!\left(\vec{I}+q\tilde{\vec{Q}}_2 \vec{\Lambda}_1\tilde{\vec{Q}}_2\rmh\right)\!\right] \notag \\\notag \\
&&=\!\mathbb{E}_{\left\{\hat{\vec{Q}}_2,\,\hat{\vec{\Lambda}}_1\right\}}\!\!\left[\log\det\!\left(\vec{I}+q\hat{\vec{Q}}_2 \hat{\vec{\Lambda}}_1\hat{\vec{Q}}_2\rmh\right)\!\right]\!\notag \\
&&=\!\mathbb{E}_{\{\hat{\vec{Q}}_2,\, \vec{W},\, \vec{Q}_1\}}\!\!\left[\log\det\!\left(\vec{I}+q\hat{\vec{Q}}_2 \vec{W}\rmh \vec{Q}_1\rmh\vec{Q}_1 \vec{W} \hat{\vec{Q}}_2\rmh\right)\!\right]\! , \;\notag\\ \eea}
\hspace{-1.6mm}where $\hat{\vec{Q}}_2$ denotes the submatrix of $\tilde{\vec{Q}}_2$ obtained by removing the last $K\!-\!M$ columns, and $\hat{\vec{\Lambda}}_1$ is the $M\!\times\!M$ submatrix by removing both the last $K\!-\!M$ rows and columns of $\vec{\Lambda}_1$. The last equality in (\ref{switch}) holds since $\vec{Q}_1\vec{Q}_1\rmh$ and $\vec{Q}_1\rmh\vec{Q}_1 = \vec{W}\hat{\vec{\Lambda}}_1\vec{W}\rmh$ have identical nonzero eigenvalues. As $\tilde{\vec{Q}}_2$ has the same distribution as $\vec{Q}_2$, the elements in $\hat{\vec{Q}}_2$ are also i.i.d. complex Gaussian distributed, and the same is true for $\hat{\vec{Q}}_2 \vec{W}\rmh$. That is to say, the ergodic capacity of the new product channel obtained by switching $K$ and $M$, i.e., the numbers of antennas of the UAV and the users in Fig.~2, is identical to the original case. Similarity, when $K\!>\!N$, one can also switch the antennas numbers of the UAV and the BTS, while the ergodic capacity remains the same. These arguments lead to a below lemma.

\begin{lemma}
The ergodic capacity of the Rayleigh product model (\ref{md2}) is invariant under permutations of the antenna parameters $(M, K, N)$. 
\end{lemma}

Following Lemma 1 and the analysis in Sec. III-A, we have Proposition 1 that states the lower-bound of the ergodic capacity for arbitrary setting of $(M, K, N)$.

\begin{proposition}
The ergodic capacity of the Rayleigh product model (\ref{md2}) is lower-bounded as
\bea \label{prop1}\tilde{S}\geq L_1\log\!\left(1+q\exp\!\big(g(K)-2\gamma\big)\right)\!,\eea
where 
 \bea g(K)=\frac{1}{L_1}\sum_{\ell=1}^{L_1}\left(\sum_{m=1}^{L_2-\ell}\frac{1}{m}+\sum_{n=1}^{L_3-\ell}\frac{1}{n}\right)\!, \notag \eea
 and $L_1\!=\!\min\{M, K, N\}$, $L_3\!=\!\max\{M, K, N\}$, and $L_2$ is the remaining element in $(M, K, N)$.
\end{proposition}

\subsection{Transmission with Optimal linear Precoding}
Next we consider the case with optimal linear precoding. That is, we assume that the UAV knows both the channel $\vec{Q}_1$ and $\vec{Q}_2$. This requires the UAVs to be more than just AF relays, since channel estimation is needed and the slot-delay $\tau$ will increase. However, we can also assume that the UAV can adjust its gesture to gradually find an optimal beamforming direction based on, e.g., measured received signal strength, and the channel estimation is not required. Nevertheless, in this section we assume that the UAV can apply an optimal linear precoder to improve the performance.

With an optimal precoding matrix $\vec{P}$, the received signal in (\ref{md2}) changes to
 \bea\label{md3}  \vec{y}&=& \sqrt{q}\vec{Q}_2\vec{P}\vec{Q}_1\vec{x}+\vec{n}. \eea
To optimize the capacity in (\ref{md3}), the precoder is set to
\bea \vec{P}\!=\!\vec{V}\rmh\vec{D}^{1/2}\vec{U}, \notag \eea
where the unitary matrices $\vec{U}$ is defined in (\ref{svd1}), and $\vec{V}$ is obtained from the SVD
 \bea \label{svd2}  \vec{Q}_2\rmh\vec{Q}_2= \vec{V}\rmh\vec{\Lambda}_2\vec{V}. \eea
The diagonal matrix $\vec{D}$ (with $d_k$ being its $k$th diagonal element) denotes the power allocation with a total-power constraint $\sum\limits_{k=1}^K d_k\!=\!K$.

With such a precoder, the capacity in (\ref{md3}) equals
{\setlength\arraycolsep{1pt} \bea \label{S4} S&=&\log\det\!\left(\vec{I}+q\vec{D}\vec{\Lambda}_1\vec{\Lambda}_2\right) \notag \\
&=&\sum_{k=1}^K\log\!\left(1+qd_k\lambda_{1}^k\lambda_{2}^k\right)\!,\eea}
\hspace{-1.4mm}where $\lambda_{1}^k$ and $\lambda_{2}^k$ are the $k$th diagonal elements of $\vec{\Lambda}_1$ and $\vec{\Lambda}_2$, respectively. The optimal $k$th diagonal element of $\vec{D}$ can be optimized through water-filling \cite{GV97}. However, evaluating the optimal ergodic capacity needs to consider joint probability distribution functions (pdfs) of $\lambda_{1}^k$ and $\lambda_{2}^k$ \cite{T99, HT04}. As we are interested in deriving a lower-bound of the ergodic capacity, to simplify the analysis\footnote{Although Marchenko-Pastur law \cite{MP67} can be used to simplify the eigenvalue distribution, it requires $K$ (as well as $M$ and $N$) to be sufficiently large, which does not hold for practical cases with a finite number of UAVs.} an equal power allocation for all transmit antennas of the UAV is assumed. That is, setting $d_k\!=\!1$ and the capacity equals
\bea  \label{S6} S=\sum_{k=1}^{K}\log\!\left(1+q \lambda_{1}^k\lambda_{2}^k\right)\!.\eea

Clearly, the number of nonzero eigenvalues in (\ref{S6}) is $L_1$, and the ergodic capacity is then lower-bounded as
\bea  \label{SL}\tilde{S}\geq L_1\mathbb{E}\!\left[\log\!\left(1+q \lambda_{1}\lambda_{2}\right)\right].\eea
where the eigenvalues $\lambda_1$ and $\lambda_2$ has the pdf \cite{T99} shown in (\ref{pdf1}) and (\ref{pdf2}), respectively, where the coefficient $\mathcal{L}_k^{n}(\lambda)$ is the associated Laguerre polynomial of order $k$ \cite{GR80}. Inserting them back to (\ref{SL}), the ergodic capacity with optimal precoding is lower bounded by the following double integral,
\begin{figure*}[b!]
\vspace{-5mm}
\hspace{0mm}\hrulefill
\vspace{1mm}
{\setlength\arraycolsep{2pt} \bea  \label{pdf1} p(\lambda_1)&=&\frac{\lambda_1^{L_2-L_1}\!\exp(-\lambda_1)}{L_1}\!\sum_{k=0}^{L_1-1}\!\!\frac{k!}{(k\!+\!L_2\!-\!L_1)!}\!\left[\mathcal{L}_k^{L_2-L_1}(\lambda_1)\right]^2\!\!.   \\
 \label{pdf2} p(\lambda_2)&=&\!\frac{\lambda_1^{L_3-L_1}\!\exp(-\lambda_2)}{L_1}\!\sum_{k=0}^{L_1-1}\!\!\frac{k!}{(k\!+\!L_3\!-\!L_1)!}\!\left[\mathcal{L}_k^{L_3-L_1}(\lambda_2)\right]^2\!\!.\eea}
 \vspace{-7mm}
\end{figure*}
\bea \label{S7} \tilde{S}\geq L_1 \int_{0}^{\infty}\!\!\!\!\int_{0}^{\infty}\!\log\!\left(1+q \lambda_{1}\lambda_{2}\right)\!p(\lambda_1)p(\lambda_2)\text{d}\lambda_1 \text{d}\lambda_2.\eea

As a special case, when $L_1\!=\!1$ it holds that $\mathcal{L}_0^{n}(\lambda)\!=\!1$, which yields a keyhole channel communication \cite{CV02, M02, AS07} with a single UAV.

\subsection{Discussions on the Parameter Designs for the UAV}
 
With the derived lower-bound, in order for $\tilde{S}\!\geq\!\tilde{R}$, it is sufficient to have (assuming $K\!\leq\!M\!\leq\!N$)
 \be K\log\!\left(1+q\exp\!\big(g(K)-2\gamma\big)\right) \geq M\log\!\left(1+pN\right )\!. \ee
That is,
 \bea \label{cond2}  q\geq \left(\!(1+pN )^{\!M/K}\!-1\right)\!\exp\!\big(2\gamma-g(K)\big). \eea
Assuming $N, M\!\gg\!K$, in which case, 
\bea   g(K)\!\approx\!\log M+ \log N+2\gamma, \notag \eea 
the condition (\ref{cond2}) becomes
  \bea  \label{cond3}  q>\frac{\exp\!\big((pMN)/K\big)-1}{MN}. \eea
Hence, for a given set of $M$ and $N$, the required transmit power for the UAV exponentially decreases in the number of antennas $K$. This makes intuitive sense according to the MIMO capacity formula \cite{T99}. Secondly, when users are in deep fading, we can assume $pMN$ is rather small, and it holds that $$\exp\!\big((pMN)/K\big)\!-\!1\approx (pMN)/K.$$ Then, the condition (\ref{cond3}) is identical to (\ref{cond1}). This is because $N$ is sufficiently large, which yields $\vec{Q}_2\rmh\vec{Q}_2\!\approx\!K\vec{I}$ according to the channel hardening and favorable propagation properties \cite{HT04, MN16} in massive MIMO systems. Therefore, the differences between the capacities $\tilde{R}$ and $\tilde{S}$ is the same as the differences in the received SNRs for these two cases.

To design such a UAV assisted communication system, it is of interest to optimize the number of antenna $K$ for a given total transmit power constraint 
\bea \label{qhat}\hat{q}\!=\!qK.\eea
Although it may not be true in practical scenarios, theoretically it is always beneficial to have more transmit antennas than to have higher transmit power per antenna under Rayleigh fading. Therefore, we consider the optimization problem to find a maximal $K$ for a given $\hat{q}$ such that the capacity-increment ratio is above a certain threshold $\eta$. That is, with (\ref{qhat}) we solve
\bea \label{K0} K_0= \argmax_{K} \;\left\{\frac{\tilde{S}(K+1)}{\tilde{S}(K)}-1\geq \eta\right\}, \eea
where $\tilde{S}$ uses the lower-bound in (\ref{SLB}) and $K$ specifies the number of antennas of the UAVs. Such an optimization is meaningful in a case that each UAV is equipped with a single-antenna, and the objective is to maximize the utility of each UAV for assisting the users.

\section{Numerical Results}

In this section, we provide simulation results to show the performance of UAV-assisted cellular communications, as well as the effectiveness of the derived lower-bound for ergodic capacity. We also elaborate on the trade-offs between the transmit power $q$ and the number of antennas $K$ used for the UAVs.

\subsection{Tightness of the Lower-Bound}
In Fig.~\ref{fig3}, we compare the ergodic capacities $\tilde{R}$ and $\tilde{S}$ for received signal models (\ref{md1}) and (\ref{md2}), respectively, and with settings $M\!=\!4$ and $N\!=\!16$. The upper-bound for $\tilde{R}$ and lower-bounds for $\tilde{S}$ with different values of $K$ are also plotted. As can be seen, the derived lower-bound for $\tilde{S}$ in (\ref{SLB}) is quite tight when $K$ is smaller than $M$. When $K$ is larger, it is also asymptotically tight as SNR increases. Furthermore, as expected, when $p$ is small, i.e., users are in deep-fading propagation, the UAV-assisted communication even with $K\!=\!1$ can provide higher capacities than a direct approach.

\subsection{Power Increment for a Small $K$}

\begin{figure}[t]
\vspace*{-2mm}
\begin{center}
\hspace*{-5.2mm}
\scalebox{0.32}{\includegraphics{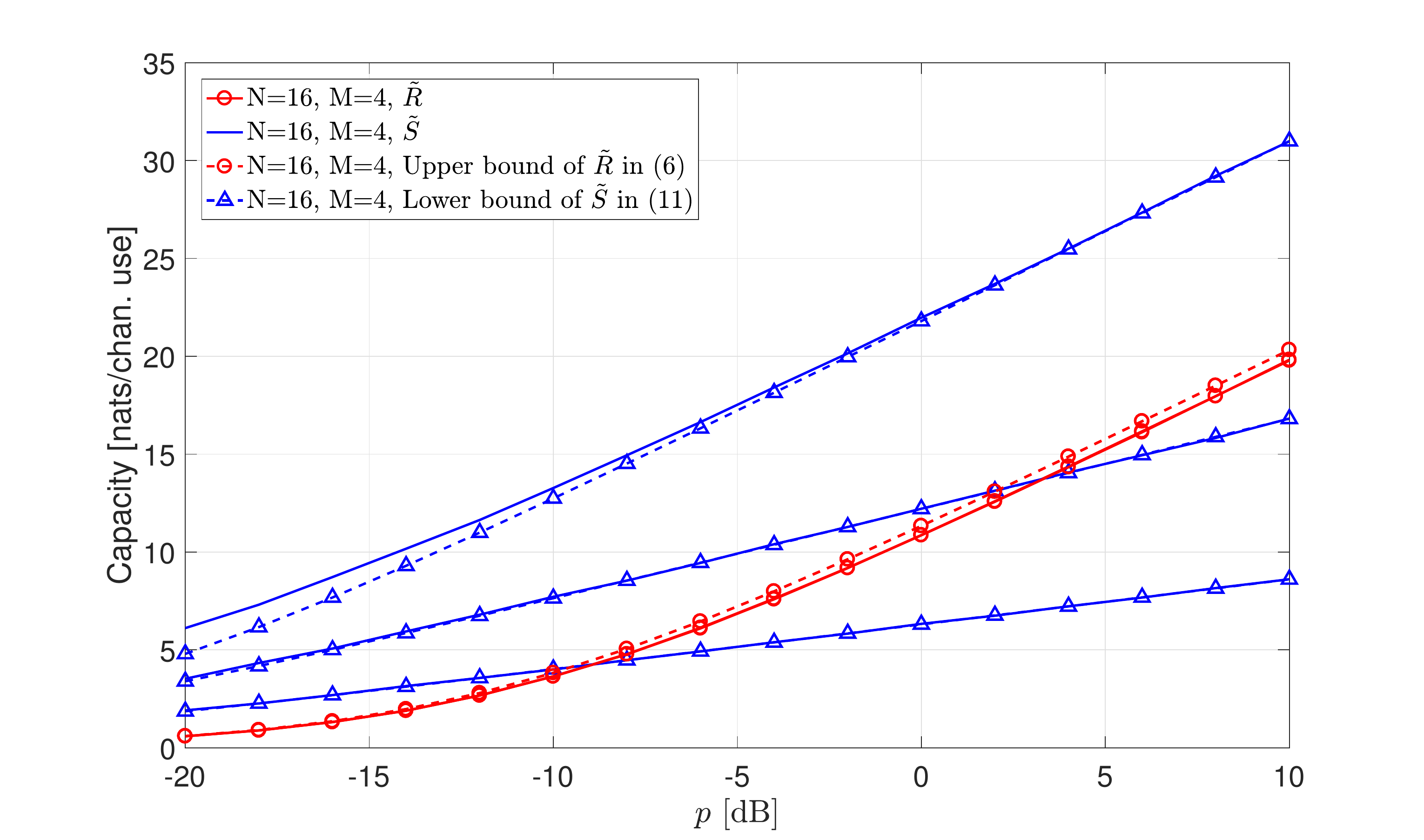}}
\vspace*{-8mm}
\caption{\label{fig3}The ergodic capacity and the derived bounds under $M\!=\!4$ and $N\!=\!16$, and with $q\!=\!10p$. From bottom to up, $K$ equals 1, 2, and 4, respectively.}
\vspace*{-4mm}
\end{center}
\end{figure}

\begin{figure}
\vspace*{-3mm}
\begin{center}
\hspace*{-5.2mm}
\scalebox{0.32}{\includegraphics{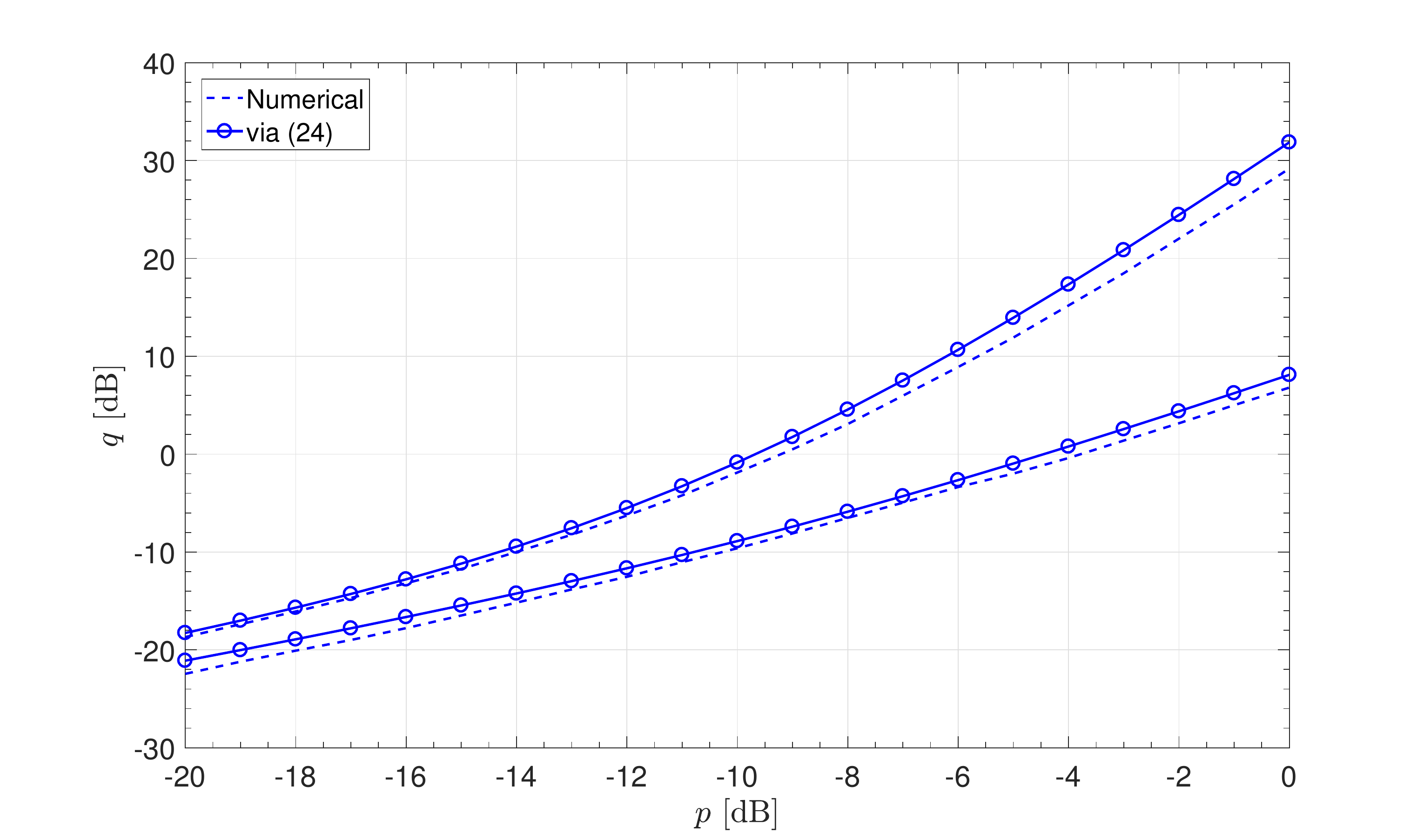}}
\vspace*{-8mm}
\caption{\label{fig4}Minimal $q$ such that the ergodic capacity $\tilde{S}\!=\!\tilde{R}$, with $M\!=\!4$ and $N\!=\!16$. From bottom to up, $K$ equals 2 and 1, respectively.}
\vspace*{-7mm}
\end{center}
\end{figure}

In Fig.~\ref{fig4}, we test the same cases in Fig.~\ref{fig3} with $K\!=\!1$ and $K\!=\!2$, and aim at finding the minimal $q$ such that the ergodic capacity $\tilde{S}\!=\!\tilde{R}$. We use with two different approaches. The first one is based on the numerical results of the ergodic capacities and the $q$ is exact. The second approach is using the derived closed-from lower-bound in (\ref{SLB}) and the value of $q$ is computed directly according to (\ref{cond2}). As can be seen, these two approaches are quite close, which validates the effectiveness of derived lower-bound.

\begin{figure}[t]
\vspace*{-2mm}
\begin{center}
\hspace*{-7mm}
\scalebox{0.32}{\includegraphics{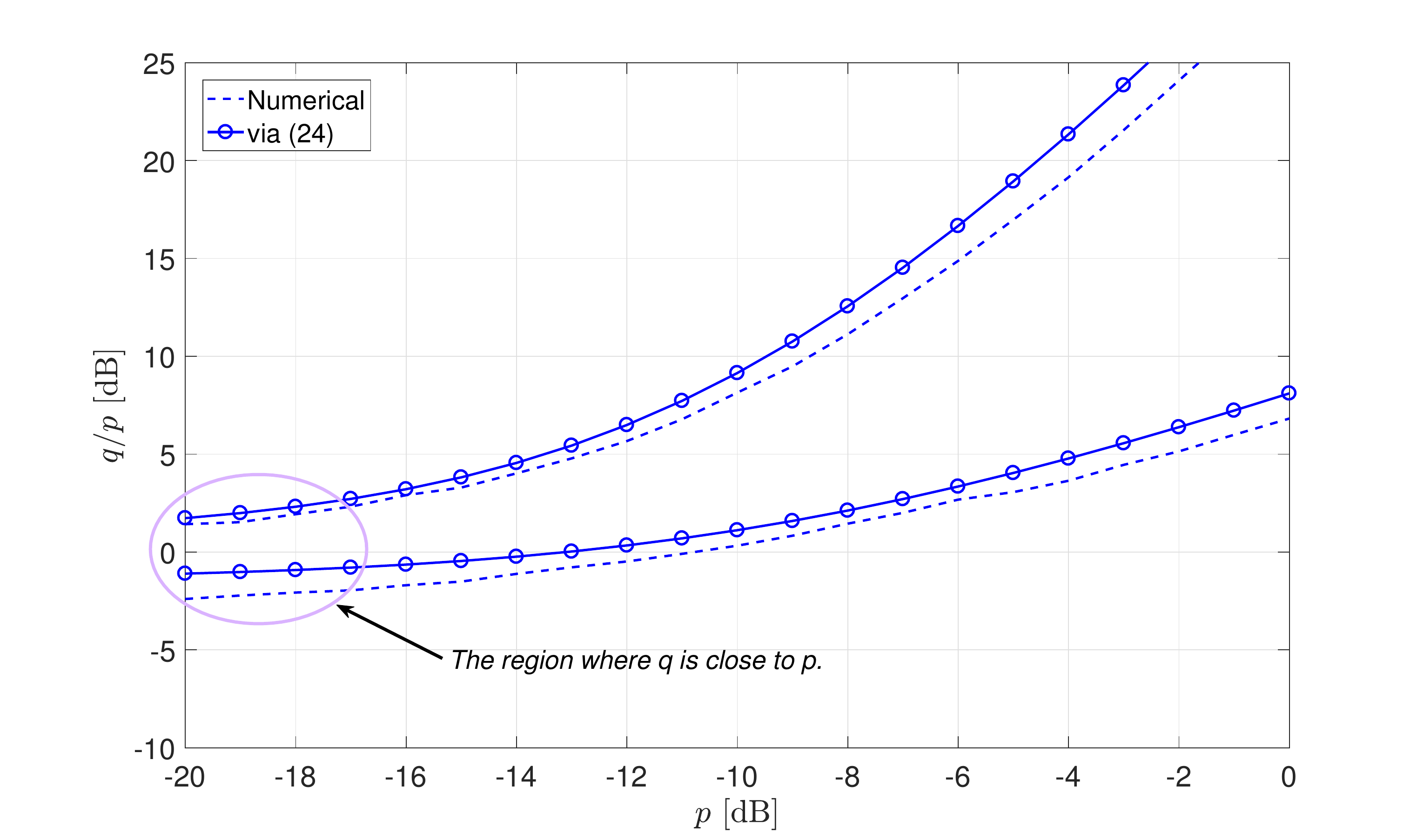}}
\vspace*{-8mm}
\caption{\label{fig7}The ratio $q/p$ of the results in Fig.~\ref{fig4}, and from bottom to up $K$ equals 2 and 1, respectively.}
\vspace*{-4mm}
\end{center}
\end{figure}

In Fig.~\ref{fig7}, we plot the ratio of $q/p$ for $K\!=\!1$ and $K\!=\!2$ in relation to $p$. An interesting observation is that when $p$ is small, the required power $q$ is even less than $p$ in order to have the same ergodic capacity. This is because the Rayleigh product channel has more degrees of freedom in the channel elements, which justifies the use of UAVs for improving the throughput of the cellular network. As predicted by (\ref{cond3}), when $p$ increases, the required power $q$ is exponentially increased in $p$, and the UAV-assisted communication becomes less power-efficient.

\subsection{Ergodic Capacity with Optimal Precoding}

\begin{figure}
\vspace*{-3mm}
\begin{center}
\hspace*{-7mm}
\scalebox{0.32}{\includegraphics{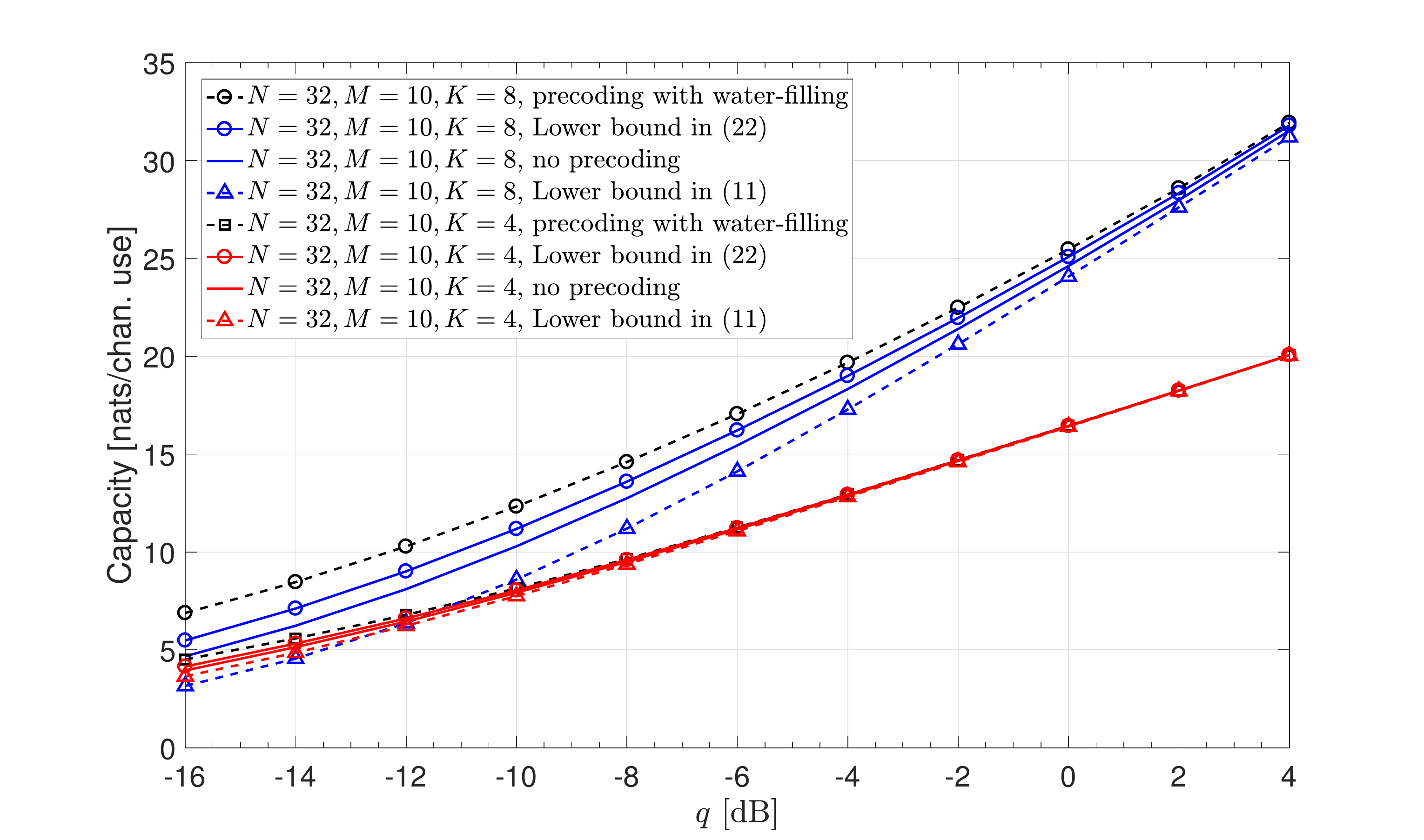}}
\vspace*{-8mm}
\caption{\label{fig5}Ergodic capacity for a fixed total transmit power $\hat{q}$ and with optimal precoder for $K$ equals 8 and 4, respectively.}
\vspace*{-7mm}
\end{center}
\end{figure}

In Fig.~\ref{fig5}, we compare the ergodic capacities $\tilde{S}$ for received signal model (\ref{md2}) with settings $M\!=\!10$ and $N\!=\!32$, and a fixed total transmit power $\hat{q}$. As can be seen, the ergodic capacity with $K\!=\!8$ provides substantial gains compared to the case $K\!=\!4$, due to higher spatial multiplexing gains. Further, with optimal precoding (based on both water-filling and equal power-allocation), the capacities are boosted in the low SNR regime. For a large $K$ or at high SNR, the gains with precoding become marginal, due to a large value of $N$. Therefore, the derived lower-bound in (\ref{SLB}) is still a good approximation for cases with linear precoding, as it is close to the lower-bound (with equal power-allocation) in (\ref{S7}) and also the optimal precoding (with water-filling).

\subsection{Trade-off Between Power and Number of Antennas}

\begin{figure}[t]
\vspace*{-2mm}
\begin{center}
\hspace*{-5.5mm}
\scalebox{0.32}{\includegraphics{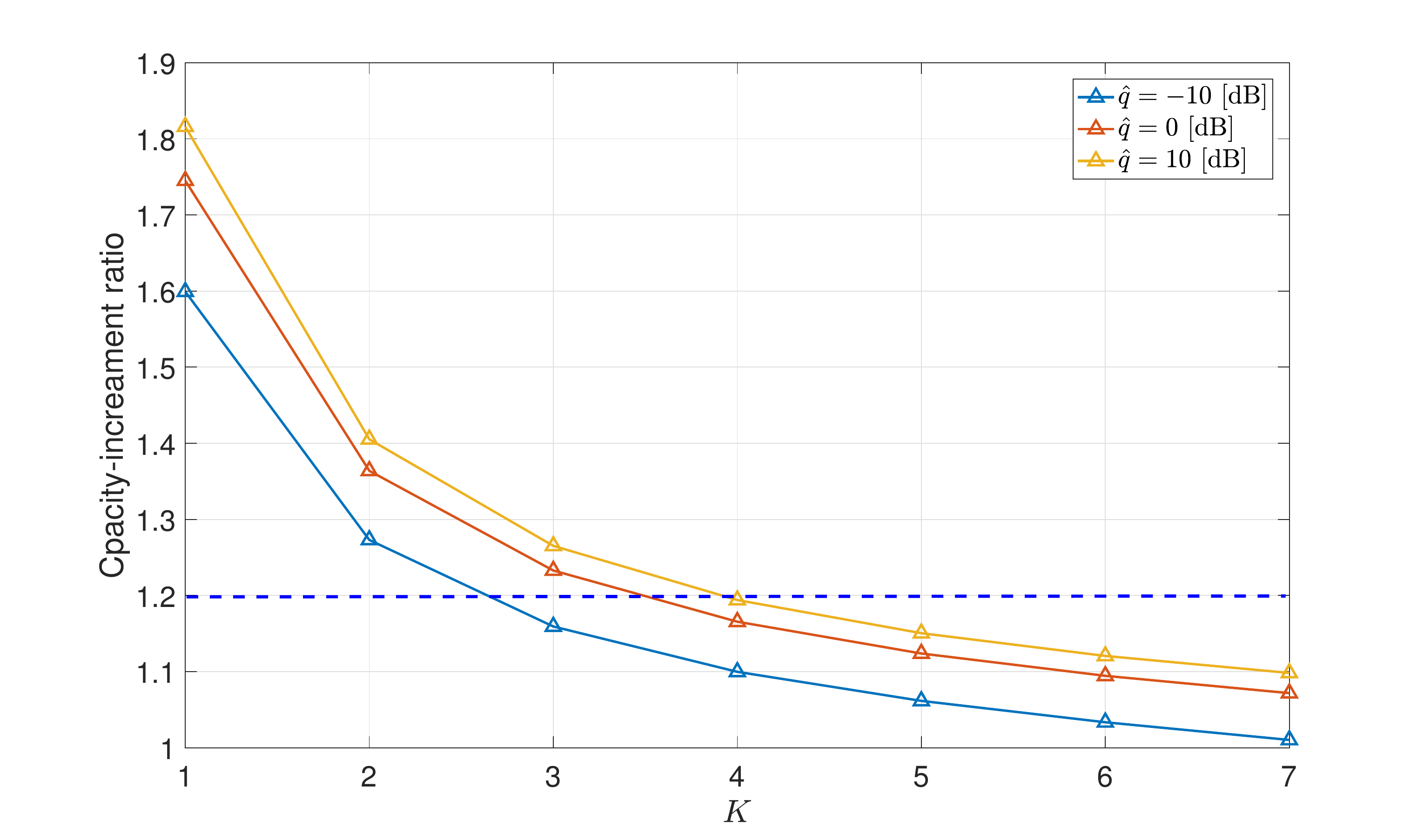}}
\vspace*{-8mm}
\caption{\label{fig6}The increment-ratio of ergodic capacity with $M\!=\!12$ and $N\!=\!32$, while $K$ increases from 1 to 8.}
\vspace*{-7mm}
\end{center}
\end{figure}

Lastly in Fig.~\ref{fig6}, we show the capacity-increment ratio with $M\!=\!12$ and $N\!=\!32$ using the derived lower-bound in (\ref{SLB}) (the numerical results are quite close and therefore not shown). If we set the utility threshold to $\eta\!=\!0.2$, the maximal values of $K$ are 3, 4, and 4 for $\hat{q}$ at -10, 0, and 10 dB, respectively. Further increasing the number of antennas (with $\hat{q}$ unchanged) will have an utility less than $\eta$. Another observation is that when $\hat{q}$ increases, the capacity increment-ratio also increases, but the gaps also gets smaller. That also means that the solution of (\ref{K0}) will converge.

\section{Summary}

We have considered an unmanned aerial vehicle (UAV) assisted cellular communication system, where the UAV is used as an amplify-and-forward relay to improve the data transmissions between a base-transceiver station (BTS) and users at cell edges or in deep shadow fading. We have modeled the channel as a Rayleigh product channel in this case, and derived a tight lower-bound of the ergodic capacity in closed-from for it. With the obtained lower-bound, analytical results has been simplified, and the behaviors of the ergodic capacity can be clearly seen in terms of the transmit power and the number of antennas of the UAV.

\end{document}